\documentstyle[12pt,epsf]{article}
\newcommand{\be}{\begin{equation}}
\newcommand{\bea}{\begin{eqnarray}}
\newcommand{\eea}{\end{eqnarray}}
\newcommand{\ba}{\begin{array}}
\newcommand{\ea}{\end{array}}
\newcommand{\ee}{\end{equation}}

\newcommand{\vm}{{\vec{m}}}
\newcommand{\vn}{{\vec{n}}}
\newcommand{\vl}{{\vec{l}}}
\makeatletter
\@addtoreset{equation}{section}

\makeatother
\expandafter\ifx\csname mathbbm\endcsname\relax

\else

\fi
\begin{document}
\baselineskip 18pt

\begin{titlepage} 
\hfill
\vbox{
    \halign{#\hfil         \cr
           SU-ITP-02/42\cr
           UOSTP-02106\cr
           hep-th/0211073 \cr
           } 
      }  
\vspace*{6mm}
\begin{center}  
{\Large {\bf 
Strong Evidence 
In Favor Of 
The Existence Of 
S-Matrix For Strings In Plane Waves}\\}
\vspace*{15mm}
\vspace*{1mm}

{\bf Dongsu Bak${}^{1,\,\,2}$ and  M. M. Sheikh-Jabbari${}^1$}

\vspace*{0.4cm}
{\it ${}^1${Department of Physics, Stanford University\\
382 via Pueblo Mall, Stanford CA 94305-4060, USA}}

\vspace*{0.2cm}

{\it ${}^2$Physics Department, University of Seoul, 
Seoul 130-743, Korea}\\

\vspace*{0.4cm}

{\tt dsbak,$\,$jabbari@itp.stanford.edu
}

\vspace*{1cm}
\end{center}

\begin{center}
{\bf\large Abstract}
\end{center}
{
Field theories on the plane wave background are 
considered. We discuss that for
such field theories one can only form $1+1$ dimensional 
freely propagating wave 
packets. We analyze tree level four point functions of scalar field 
theory as well as scalars coupled to gauge fields in detail 
and show that these 
four point
functions are well-behaved so that they can be interpreted 
as S-matrix elements 
for 2 particle $\to$ 2 particle scattering amplitudes. 
Therefore, at least 
classically, field theories on the plane wave background  have S-matrix 
formulation.
}

\end{titlepage}

\vskip 1cm

\section{Introduction}

As a usual lore, our understanding of (perturbative) string theory 
is based on 
the S-matrix interpretation of the string scattering amplitudes, given as
correlators of vertex operators and computed using the worldsheet conformal 
field theory.
However, generally the question whether a field/string theory
has an S-matrix, at least at classical level, returns to the question of  
having well-defined asymptotic
states, i.e. freely propagating  states in asymptotic regions of spacetime.
For any field theory with a well-defined (positive definite) Hamiltonian,
this is equivalent to having a potential 
that has some flat directions. However, having these asymptotic flat 
directions is not sufficient and one should further check 
that the potential at large separation
falls off fast enough. In the field theory terminology and in terms 
of Feynman diagrams, this means that we should have 
well-behaved
tree level scattering amplitudes.
Then one can show that any Lorentz invariant local $D$ ($D>2$) dimensional 
field theory on flat background has S-matrix.

The question about the existence of S-matrix 
becomes more non-trivial for the case of field/string
theories on non-flat spaces. A famous example is 
the string theory
on 
AdS space, where despite of having well-defined correlators,
these
correlators do not have an S-matrix interpretation. This is also manifest
in the dual gauge theory in the fact that, it is conformally invariant.
However in the AdS case, taking the large AdS radius limit one can show that
the correlators indeed recover the S-matrix elements of flat space field
theories.

Here we study the  existence of the S-matrix for string/M theory on
the plane wave background. The plane wave geometries are appearing as a 
Penrose 
limit \cite{Penrose} of $AdS_p\times S^q$ spaces and hence it is natural to 
expect 
strings on plane waves to have a description in terms of a specific subsector 
of the operators of ${\cal N}=4$ SYM on $R\times S^3$, namely the large 
$R$-charge sector \cite{BMN} (which hereafter will be called BMN sector).
Since the appearance of the paper by Berenstein-Maldacena-Nastase
(BMN) \cite{BMN}, there have been many 
papers
trying to check the BMN conjecture. In these works mainly two directions 
have been pursued. Some have  focused on the gauge theory 
calculations such as \cite{{SYM},{mixing1},{mixing2}} and some are devoted to the string 
theory side 
\cite{{SFT1},{SFT2}}, with the aim of reproducing the gauge theory results 
through string field theory  amplitudes. 

However, the formulation of string vertex operators in the plane wave 
background, as the conventional tool for doing string scatterings, has not 
been developed yet. As the first step to make sure that such vertex operators 
exist, 
one must have an  affirmative answer to the question of existence of S-matrix 
for strings in plane wave background.
One may try to argue that, since the plane wave
geometry is coming as a Penrose limit of the AdS, similar to its AdS parent,
we do not have S-matrix in this case. This may be further supported by the
fact that string theory on plane wave is dual to the BMN (large $R$-charge) 
sector of a conformal gauge theory and hence the question of the existence of
the S-matrix has the same answer as the AdS background. However, on the other
hand, the
Penrose limit is in fact a particular way of taking the large radius limit  
in AdS, where it is generally believed that one has 
a well-defined S-matrix.
Moreover, since under the Penrose limit 
the whole causal structure as well as asymptotic behavior of the AdS is 
changed (e.g. now we have a one
dimensional, light-like boundary) \cite{causal}, it is not so obvious that 
the answer to the question of existence of
S-matrix follows directly from the AdS parent.

Specifically here we consider field theories on a $D$ dimensional plane
wave geometry:
\be\label{wave}
ds^2=-2dx^+dx^--\mu^2 \sum_{i=1}^{D-2} z_i^2 (dx^+)^2+ \sum_{i=1}^{D-2}     
dz_idz_i\ .
\ee
The plane wave geometry produces confining potentials in the
transverse directions. The shape of the potential is quadratic in the
transverse coordinates for massless as well as massive particles and thus
it is certainly impossible to propagate freely in the transverse
directions. (In other words there is no translation symmetry along $z_i$ 
directions.) Then, the only potentially flat direction is $x^-$ and what we
mean by the S-matrix formulation should
be understood in this $1+1$ dimensional sense, with $x^-$ as the  spatial 
dimension.
Hence the nature of interaction along this one spatial dimension
will be the focus of our analysis in this work.

It is well-known that effective dynamics of strings is 
governed by field theories, in particular effective dynamics of strings
in a plane wave background is governed by supergravity in the same background
(for an explicit form of such an action  for the axion-dilaton field of IIB 
supergravity see e.g. \cite{Kim, MT}). 
Therefore, here instead of strings on plane 
waves we consider field theories on the plane wave 
background and study the existence of S-matrix for those field theories. 
Since we are dealing with effective two dimensional theories, 
at first sight, it 
may seem that we are going to face the usual problem of 
two dimensional massless 
theories, namely Green's function does not fall off (in fact it grows) at 
infinity. However, 
we are in fact safe 
because of the leakage of the wave-function 
to the ``transverse'' directions. In particular 
we note that the 
nature of the leakage is so that the less 
the exchanged  light-cone momentum, the more 
possibility for the wave-function to spread out 
in the transverse directions. 
This is one of the essential mechanisms for the existence of 
the S-matrix. In the following, 
through explicit field theory computations, 
we show that generic tree level 
amplitude for such field theories is  well-behaved so that 
consequently these field theories admit  S-matrix formulations.

Organization of the paper is as follows. In Section 2 we consider scalar field 
theory with $\phi^3$ and $\phi^4$ 
interactions on the $D$ dimensional plane wave 
background. Calculating the $2\phi\to 2\phi$ scattering amplitude we show that 
these amplitudes are smooth enough to have well-defined S-matrix description. 
Because of the form of the plane wave geometry (\ref{wave}) it turns out to be 
more convenient to use the light front coordinates. Due to peculiar features 
and potential problems of field theories in the light-cone gauge (in the $p_- 
\to 0 $ limit) \cite{Lenny}, higher spin fields should be dealt with 
separately. Hence in 
Section 3 we consider scalars coupled to vector (gauge) fields. Studying 
the $2\phi\to 2\phi$ amplitudes mediated by the gauge fields,
we show that the gauge theories     
 on plane wave background also admit S-matrix description, the result which we 
believe to be true for any generic higher spin theories including 
gravity. The last section contains our conclusions and remarks. In Appendix A, 
we have gathered some identities used in the computations and in Appendix B,
through the most general two body non-relativistic quantum mechanical 
arguments, we have made clear how the exchanged momentum 
$p_- \to 0$ corresponds to the large $x^-$ 
separation.   

\section{Scalar Field Theories on Plane Wave Background}

We first consider scalar  theories on the plane wave background, which would 
eventually reveal the essence of the question posed. 
With the standard kinetic term, we consider 
only the local interactions. 
The interactions are then limited to the forms of local products.
We focus on the nature of  tree level two particle 
interactions from the view point of 1+1 dimensions. Considering
cubic interactions, one may build up the four point tree amplitudes
corresponding to two body potentials. Thus looking at the behavior
of the four point interaction, one can check  whether  free
states in the large $x^-$ separation are allowed or not.
As we shall see below, these derived interactions fall off fast enough
at large separation, so that the asymptotic free states indeed exist.
The local quartic scalar interaction contributes to the four point 
amplitudes but, again, 
does not affect the existence of the asymptotically
free states. The interactions of higher  than fourth order
do not contributes four point tree amplitude.

\subsection{The model} 
Consider a scalar field theory on the  $D$ dimensional plane 
wave geometry (\ref{wave}) with the usual conventional kinetic terms: 
\be\label{Lagrangian}
{\cal L}=\partial_+\phi\partial_-\phi-{1\over 2} 
\mu^2 \sum_{i=1}^{D-2} z_i^2 (\partial_-\phi)^2-{1\over 
2}(\partial_i\phi)^2-{1\over 2}M^2\phi^2-{\lambda\over 
3!}\phi^3\,,
\ee
where we have added a mass term, which is not typically there for a
supergravity field in the plane wave background, as well as a 
cubic interaction term. As we can see explicitly from the Lagrangian, for a 
fixed 
$\mu$, the above action is not invariant under Lorentz boost in the 
$x^{\pm}$ plane.  Nonetheless it is quite natural to use the light-cone 
frame. In the following we take $\mu$ to be positive without loss of
generality. 
The classical equation of motion for the free  theory reads
\be\label{eom}
-2\partial_+\partial_-\phi+\mu^2  
\sum_{i=1}^{D-2} z_i^2 \partial_-^2\phi+\partial_i^2\phi-M^2\phi=0\ .
\ee

The $z_i$ dependence of solutions to the above equation is a $D-2$ dimensional 
harmonic oscillator wave function with frequency $\mu p_-$, which we will 
denote it by $K_{n_i},\ i=1,2,..., D-2$.   Thus, 
\be\label{Phi}
\phi=
\sum_{\vn}{1\over 2\pi}\int {dp_+}
\int_0^\infty {\,\,\,dp_- \over 
\sqrt {2 p_-}}\,\,
 \phi_{\vn}(p_-,p_+)\prod_{i=1}^{D-2} K_{n_i}({\sqrt{\mu p_-}z_i}) 
e^{-i(p_-x^-+p_+ x^+)}+ {\rm h.\ c.} 
\ee
solves the above equation if
\bea\label{dispersion} 
&& p_+=\mu\ {\cal N}+{M^2\over  2p_-}\equiv {\cal E}(\vn,p_-)\ ,\nonumber\\
&& {\cal N}=(\sum_{i=1}^{D-2} n_i)+ {D\!-\!2\over 2}\ ,
\eea
with $\{n_i\}=\vn$. Here $p_+$ may be 
thought as the light-cone Hamiltonian of the  above field theory.\footnote
{It is well-known that a boost along $x^-$ direction is in fact a 
scale transformation along the longitudinal direction, i.e.
\[ x^-\to e^{v}x^-
\]
where $v$ is the hyperbolic boost angle and
\[ p_-\to e^{-v}p_-, \ \ \ \ \ \ n_i\to n_i\ .
\]
In the light-cone frame for the flat space, the
light-cone Hamiltonian transforms 
as
$p_+\to e^{v}p_+.$ However, in our case $p_+$ would have the usual behavior 
only if one also transforms $\mu \to e^v \mu$.
 Hence for a fixed $\mu$ the two 
dimensional boost invariance is lost, and for the same reason the creation 
operators $\phi_{\vn}$ have not a simple transformation under boost. 
It is important to note that $p_-\to 0$ region, unlike the usual flat space 
light-cone frame, cannot be studied through a simple longitudinal boost.}
Since we are in the light-cone frame, all the $p_-$ momenta are positive
definite. This 
has been made manifest in the Eq.(\ref{Phi}). (As usual $p_-=0$ in the 
light-cone is 
problematic and we shall not consider these complications.)
 
Because of the harmonic oscillator potential in the transverse 
directions ($z_i$ dependent part), fields have a 
Gaussian fall-off and hence we 
do not have the asymptotic free states for those directions. 
However, the theory may 
effectively be treated as a two dimensional theory of component 
fields $\phi_{\vn}$, 
having masses greater than or equal to $ {D-2\over 2}\mu$. 
Therefore, the question about S-matrix is reduced to the 
question about 
the two dimensional effective field theory. To study the theory it would be 
helpful to rewrite  Eq.(\ref{Lagrangian}) in terms of the $\phi_{\vn}$ 
modes.
Using the orthonormality of the $K_{n_i}$ functions and integrating 
over $z_i$ coordinates, the effective two dimensional Lagrangian 
becomes 
\be\label{2dimLag}
{\cal L}_2= \sum_{\vn} 
\phi^*_{\vn}(p_-,p_+)\left( p_+ 
-
{\cal E}(\vn,p_-)
\right) \phi_{\vn}(p_-,p_+)\ +{\cal L}_{int}\ ,
\ee
with
\bea\label{Lint}
&& {\cal L}_{int}=-
\lambda \sum_{\{\vn,\vm,\vl\}}
{1\over (p_-^1p_-^2p_-^3)^{1/2}}\phi_{\vn}(p_-^1,p_+^1)
\phi_{\vm}(p_-^2,p_+^2)\phi^*_{\vl}(p_-^3,p_+^3) 
\nonumber\\ 
&&\ \ \ \ \ \ \ \ \ \ \ \ \times \,\,
F_{\{\vn,\vm;\vl\}}(p_-^1,p_-^2;p_-^3)\ \delta(p_-^1+p_-^2-p_-^3)
\delta(p_+^1+p_+^2-p_+^3)+ {\rm h.\ c.} ,
\eea
where $F$ is defined by
\bea\label{Fs}
F_{\{\vn_1,\vn_2;\vn_3\}}(p_-^1,p_-^2;p^3_-)&=&
\prod_{i=1}^{D-2}\int dz_i\,
K_{n_i^1}\left(\sqrt{\mu p_-^1}z_i\right)K_{n_i^2}\left(\sqrt{\mu p_-^2}
z_i\right)
K_{n^3_i}\left(\sqrt{\mu 
p^3_-}z_i\right)\cr
&=& \int d\vec{z}\,\, K_{\vn_1} K_{\vn_2}K_{\vn_3}\,.
\eea
For the notational simplicity, we have introduced $K_\vn\equiv 
\prod_{i=1}^{D-2}K_{n_i}(\sqrt{\mu p_-}z_i)$. 

In this model, we 
have conserved $p_-,p_+$ and $SO(D-2)$ quantum numbers at each 
vertex.  
(Note that in general $\{\vn\}$ are not conserved.) 
The conservation of $p_\pm$ is explicit due to
the presence of the delta functions while the $SO(D-2)$
conservation is not and 
follows from the properties of $F_{\{\vn_1,\vn_2;\vn_3\}}$.
One should note that as it is explicit in Eqs.(\ref{Lint}) and (\ref{Fs}) and 
due to a non-trivial 
dependence of $F$'s on all the momenta, although the $D$ dimensional theory is 
a local field theory, the effective theory of $\phi_{\vn}$ modes as a two 
dimensional field theory is non-local.
Note again that  the momentum $p_-$ takes only positive definite
values while $p_+$ may be any real number.

\begin{figure}[htb]
\vskip 1cm
\epsfxsize=3.8in
\centerline{
\epsffile{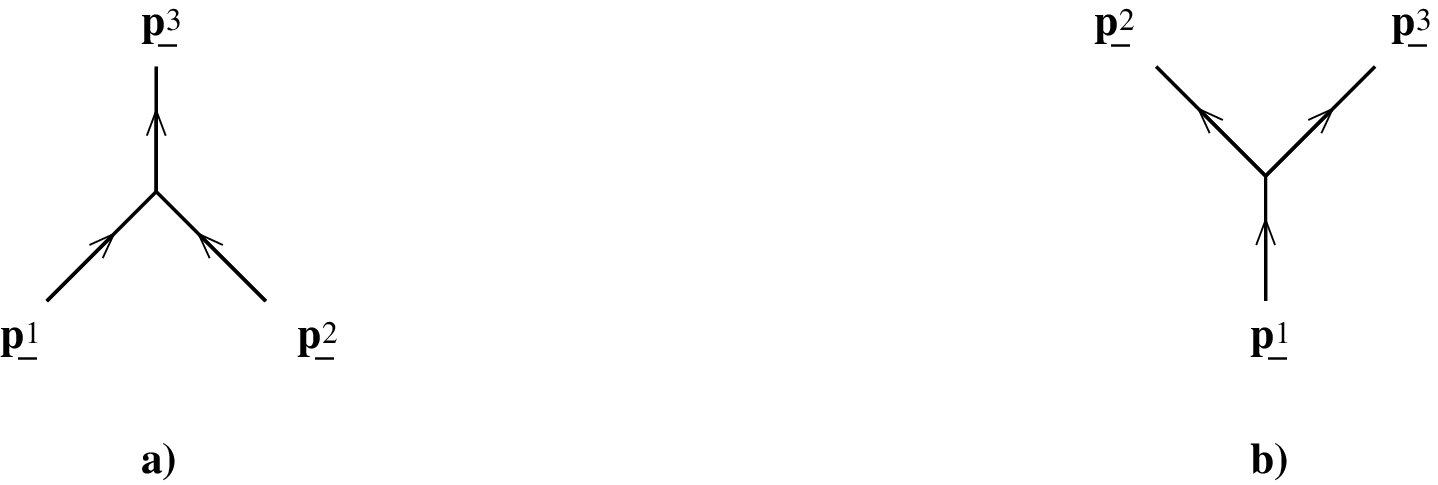}
}
\vspace{.1in}
{\small Figure~1: 
For the vertex a), the coupling is given by
$\lambda 
(p_-^1p_-^2p_-^3)^{-1/2}
F_{\{\vn_1,\vn_2;\vn_3\}}(p_-^1,p_-^2;p^3_-)$; for the vertex b)
one has 
$\lambda 
(p_-^1p_-^2p_-^3)^{-1/2}
F_{\{\vn_2,\vn_3;\vn_1\}}(p_-^2,p_-^3;p^1_-)$. There are only two kinds of
 cubic 
vertices in the scalar theory.}
\vskip 0.5cm 
\end{figure}

\subsection{Evaluating the scattering amplitude}

Now to check whether the theory has an S-matrix or not, one should check if 
there
are long-range forces in the system and study the effective forces between 
particles at large $x^-$ separation. 
In order that, let us focus on the simplest tree level scattering  
of on-shell particles and begin  with the assumption that we have  
well-defined asymptotic free particle states. Explicitly, we consider the 
$$
\phi_{\vn_1}\phi_{\vn_2}\to \phi_{\vn_3} \phi_{\vn_4}
$$
scattering at tree level.

\begin{figure}[htb]
\vskip 1cm
\epsfxsize=5.0in
\centerline{
\epsffile{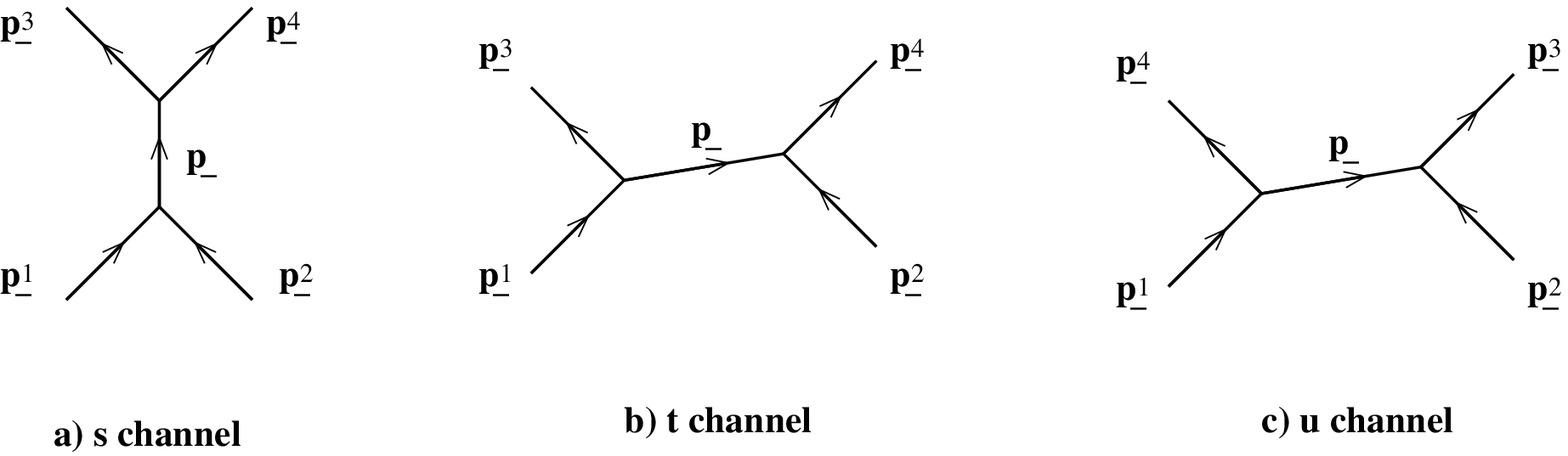}
}
\vspace{.1in}
{\small Figure~2: Four point amplitudes of the scalar theory.
The $t$ and $u$ -channel diagrams are for $p_-^1\ge  p_-^3$ and $p_-^1\ge  
p_-^4$, respectively. }
\vskip 0.5cm
\end{figure}

The  propagator 
of $ \phi_{\vn}$ fields reads
\be\label{2ptfun}
\langle\  \phi_{\vn}(p_-,p_+) \phi^\dagger_{\vm}(p'_-,p'_+)\ \rangle
={i\over p_+ -
{\cal E}(\vn, p_-)
+i\epsilon}
\ \delta_{\vn,\vm}
\delta(p_+-p'_+)\delta(p_--p'_-)\ .
\ee
Due to the nonrelativistic nature of the propagator, the amplitudes
are non-vanishing only when all the propagators are forwarded in 
$x^+$ direction.
The two particle scattering amplitude can be in $s$,  $t$ or 
$u$ -channels. The amplitude for $s$-channel is
\be\label{Samp}
{\cal A}^s_{2\to 2}\sim {\lambda^2\over (p_-^1p_-^2p_-^3p_-^4)^{1/2}} 
\sum_{\vn} {
F_{\{\vn_1,\vn_2;\vn\}}(p_-^1,p^2_-;p_-) 
F_{\{\vn_3,\vn_4;\vn\}}(p_-^3,p^4_-;p_-)
\over p_-p_+ - 
p_- {\cal E}(\vn, p_-)
}
\ ,
\ee
where the sum is over the allowed set of ${n_i}$ states,
\bea
p_-^1+p_-^2&=&p_-^3+p_-^4=p_-\ , \cr
p_+^1+p_+^2&=&p_+^3+p_+^4=p_+\,. 
\eea
As for $t$-channel, 
\be\label{Tamp}
{\cal A}^t_{2\to 2}\sim {\lambda^2 \over (p_-^1p_-^2p_-^3p_-^4)^{1/2}
} 
\sum_{\vn} {
F_{\{\vn,\vn_3;\vn_1\}}(p_-,p_-^3;p_-^1) 
F_{\{\vn,\vn_2;\vn_4\}}(p_-,p_-^2;p_-^4)
\over p_-p_+ - p_-
{\cal E}(\vn, p_-)}
\ ,
\ee
with
\bea
p_-^1-p_-^3&=&p_-^4-p_-^2=p_- \ ,\cr
p_+^1-p_+^3&=&p_+^4-p_+^2=p_+ \,.
\eea

In order to have a well-defined S-matrix it is necessary (though not 
sufficient) the above 
amplitudes for generic values of the parameters of external on-shell 
particles to be  finite. For this the 
following two conditions should be met:
\newline
{\it i)} the propagator for the exchanged particle should never blow up, i.e.
the exchanged particle should always be off-shell,\footnote{
Of course one should note that generally when we have unstable particles it is 
quite possible that in a 3-particle vertex we have all three particles 
on-shell simultaneously. The famous example of this is the decay of Z or Higgs 
boson into two fermions. However, because of quantum corrections to the 
propagator of the unstable particle (which physically, through 
non-relativistic Wigner-Breit formula, correspond to the width of the 
particle) in fact the 
propagator does not blow up when the particle is on-shell. 
(For a more detailed arguments on this issue see e.g. \cite{Peskin}.)
 In our cases, however, 
the field theories are coming as effective theories of 
string theory and hence in our field theory analysis the quantum corrections are 
not an issue. Moreover, 
we expect our particles at least for $M=0$ case (corresponding to SUGRA 
modes) to be stable.}    
and \newline
{\it ii)} the sum over all the possible excitations of the exchanged particle
should be convergent.

In what follows we shall check explicitly that the above two conditions are 
indeed  satisfied.

\subsubsection{Exchanged particle is never on-shell}

As we know in the usual scalar field theories
as a direct result of Lorentz invariance and momentum 
conservation at each vertex, all the particles appearing at a vertex cannot be
on-shell simultaneously. However in our problem, due to the lack of Lorentz 
invariance and the specific spectrum of our problem, it is not clear that this 
is impossible. In fact, if we only focus on the momentum conservation and 
put two of the particles on-shell there is nothing to prevent us from 
having the third particle on-shell, as well. Explicitly,
let us consider the ``massless'' case, i.e. $M^2=0$ and 
in-going on-shell particles to have $(p_{\pm}^1, \vn_1), (p_{\pm}^2, 
\vn_2)$:
\bea
p_+^1&=&\mu (\sum n_i^1+{D-2\over 2})\ , \cr
p_+^2&=&\mu (\sum n_i^2+{D-2\over 2})\ . 
\eea
Then, in the $s$-channel, the exchanged particle would have
\be\label{onshell}
p_+=p_+^1+p_+^2=\mu \left(\sum (n_i^1+n_i^2)+D-2\right)\ .
\ee
Hence, for any {\it even} $D$, $p_+/\mu$ can 
be written as $\sum l_i+{D-2\over 
2}$. One may easily check that this argument appears also 
working for $t$ or $u$ 
channels.

However,  to know whether 
the on-shell  propagation for the internal lines 
is actually occurring, one should check 
whether they really appear in the  amplitudes. In other words, 
one should check if the 
corresponding $F_{\vn_1,\vn_2;\vn}$ functions vanish or not. 
For that we need to work out 
explicit form of the integrals of Eq.(\ref{Samp}). This has been done in the 
Appendix A, and the results for the $s$-channel  is that
\be\label{sf}
F_{\vn_1,\vn_2;\vn}=0 \ \ \ \ {\rm for}\ \ \ n_i^1+n_i^2 < n_i\,.
\ee
For the internal line in this case, one has
\be
p_+ -{\cal E}(n_i, p_-)=  \mu \sum_i (n^1_i+
n_i^2-n_i) +\mu {D-2\over 2} +{M^2\over 2}\left({1\over p^1_-}+{1\over p^2_-}
-{1\over p_-}
\right) 
\ee
When $F$ is non-vanishing, i.e. $n_i^1+n_i^2 \ge n_i$,
the denominator, $p_+ -{\cal E}(\vn, p_-)$, 
is clearly positive and non-zero. 
Similarly for the $t$-channel,
\be
p_+ -{\cal E}(\vn, p_-)=  \mu \sum_i (n^1_i-
n_i^3-n_i) -\mu {D-2\over 2} +{M^2\over 2}\left({1\over p^1_-}-{1\over p^3_-}
-{1\over p_-}
\right) 
\ee
with $p^1_-=p^3_-+p_-$. Again from the Appendix A, one has 
\be\label{sf1}
F_{\vn,\vn_3;\vn_1}=0 \ \ \ \ {\rm for}\ \ \ n_i^1>n_i^3 + n_i\,,
\ee
and for  non-vanishing $F$, 
$p_+ -{\cal E}(n_i, p_-)$ is negative definite.
Therefore, the on-shell condition and the non-zero $F$ condition cannot be 
satisfied simultaneously.  

\subsubsection{The sum over internal excitations is convergent}

In order to check that the sum over the possible internal excitations is 
convergent, and hence the scattering amplitude is finite,   
we need to have $F$'s.  
In general $F$'s are momentum dependent 
and hence they are different for $s$, $t$ or $u$ channels of
the four point amplitudes. Let us first consider 
the $s$-channel contributions to the  scattering amplitude.
Using Eq.(\ref{3hermit}) we find
\bea
&& F_{\vn_1,\vn_2;\vn}(p_-^1,p_-^2;p_-)= 
\left({\mu p_-^1\mu p_-^2 \over \pi \mu 
p_-}\right)^{{D-2\over 4}} \prod_{i=1}^{D-2}\Biggl[ {1\over 
\sqrt{2^{n_i^1+n_i^2+n_i} n_i^1!n_i^2! n_i!}} \cr 
&&\ \ \ \ \ \ \ \ \ \ \ \ \ \ \ \ \ \ \times
\sum_{l^1_i=0}^{n_i^1}\sum_{l^2_i=0}^{n_i^2} A_{l^1_i n^1_i}(\alpha) 
A_{l^2_i n^2_i}(\beta) 
{2^{s_i} l^1_i!l^2_i!n_i!\over (s_i-n^1_i)!(s_i-l^1_i)!(s_i-l^2_i)!}\Biggr]
\eea
where $2s_i=l^1_i+l^2_i+n_i$, 
$s_i\geq max\{n_i,l^1_i,l^2_i\}$, $\alpha^2={p_-^1\over 
p_-}$,  $\beta^2={p_-^2\over p_-}$, and
$A_{l_i n_i}(\alpha)$ functions 
are given in Eq.(\ref{a_k}). Because of momentum conservation 
$\alpha^2+\beta^2=1$.  The other vertex factor 
$F_{\vn_3,\vn_4;\vn}(p_-^3,p_-^4;p_-)$
has the same expansions once  $p^1_-,p^2_-$ and $\vn_1,\vn_2$ are 
respectively replaced with $p^3_-,p^4_-$ and $\vn_3,\vn_4$.
We observe the following two properties
of above $F$ functions. First  they are finite  for any finite
external momenta because $p_- = p^1_-+p^2_-=p^3_-+p^4_-$. 
Second they vanish
whenever $n_i > n^1_i + n^2_i$ or $n_i > n^3_i + n^4_i$. 
Hence the range of sum on $n_i$ in Eq.(\ref{Samp}) 
is always bounded from above in the amplitudes with
each term finite. This implies that the $s$-channel amplitudes
are always finite once external momenta are non-vanishing.

For the $t$-channel we would have similar expressions for $F$; 
explicitly they are given by
\bea
&& F_{\vn,\vn_3;\vn_1}(p_-,p_-^3;p_-^1)= 
\left({\mu p_-\mu p_-^3 \over \pi \mu 
p^1_-}\right)^{{D-2\over 4}} \prod_{i=1}^{D-2}\Biggl[ {1\over 
\sqrt{2^{n_i^1+n_i^3+n_i} n_i^1!n_i^3! n_i!}} \cr 
&&\ \ \ \ \ \ \ \ \ \ \ \ \ \ \ \ \ \ \times
\sum_{l^3_i=0}^{n_i^3}\sum_{l_i=0}^{n_i} A_{l^3_i n^3_i}(\alpha) 
A_{l_i n_i}(\beta) 
{2^{s_i} l^3_i!l_i!n^1_i!\over (s_i-n^1_i)!(s_i-l^3_i)!(s_i-l_i)!}\Biggr]
\eea
where $2s_i=n^1_i+l_i+l^3_i$, 
$s_i\geq max\{n^1_i,l_i,l^3_i\}$, $\alpha^2={p_-^3\over 
p^1_-}$, $\beta^2={p_-\over p^1_-}$ and again
$\alpha^2+\beta^2=1$ due to the momentum conservation.
The other vertex factor 
$F_{\vn,\vn_2;\vn_4}(p_-,p_-^2;p_-^4)$
has the same expansions as the above 
once  $p^1_-,p^3_-$ and $\vn_1,\vn_3$ are 
replaced with $p^4_-,p^2_-$ and $\vn_4,\vn_2$, respectively.

The terms in $F$ do not in general vanish for arbitrarily 
large $n_i$.  
However one can show that the sum over $n_i$ 
converges, leading to finite results.
Besides convergence of the sum, one should also check the
behaviors of the amplitude when $p_-$ approaches to zero. In this region,
one can prove that the most singular contribution
comes as 
\bea
{\cal A}^t_{2\to 2}\sim \left\{
\begin{array}{ll}
{\lambda^2\over \sqrt{p_-}}\times {\rm finite\ part}\ \ \ \ \ \ \ \ \ \ 
& {\rm for \ D=3}\\
\,\,\lambda^2\times {\rm finite\ part} \  &{\rm for \ D> 3}
\end{array}
\right.
\label{singular} 
\eea
The corresponding interactions between two particles in the 
asymptotic region is still weak enough to allow us to define 
free asymptotic states as we shall see below.

To show such  properties of $F$, let us consider the case where
$\vn_1$ and $\vn_3$ are even with $\vn_1=2\vm_1$
and $\vn_3=2\vm_3$. Then $\vn$ has to be even  to have a non-zero
$F$, which we denote by $\vn= 2 \vm$. Then, one  has
\bea
&& F_{2\vm,2\vm_3;2\vm_1}(p_-,p_-^3;p_-^1)= 
\left({\mu p_-\mu p_-^3 \over \pi \mu 
p^1_-}\right)^{{D-2\over 4}} \prod_{i=1}^{D-2}\Biggl[ 
{\sqrt{2m_i!} \sqrt{2m^1_i!} \sqrt{2m^3_i!} 2^{-m_i}} \cr 
&&\ \ \ \   \times
\sum_{l^3_i=0}^{m_i^3}\sum_{l_i=0}^{m_i} 
{(-1)^{m_i-l_i+ m^3_i-l^3_i}
(\beta^{2})^{l_i +m^3_i -l^3_i} (1-\beta^2)^{l^3_i+m_i-l_i}
2^{l_i+l^3_i-m^3_i}
\over (m_i-l_i)!(m^3_i-l^3_i)!
(m^1_i + l^3_i- l_i)! (l^3_i + l_i- m^1_i)!(m^1_i + l_i- l^3_i)!}
\Biggr]
\label{eventchannel}
\eea
with extra condition on the sum over $l_i$ by
$|m^1_i - l^3_i|  \le l_i \le m^1_i + l^3_i$. 
From the expression one sees that
each individual term is finite and behaves as
$(p_-)^{{D-2\over 4} + \sum_i (l_i +m^3_i-l^3_i)}$, which
 goes to zero as one sends $p_-$ to zero.
In particular the lowest power term has a form
$(p_-)^{{D-2\over 4} + \sum_i |m^1_i-m^3_i|}$
(since $l_i \ge |m^1_i-m^3_i|$).

Considering generic terms appearing in the ${\cal A}^t$ amplitudes,
one finds that the term having lowest power of $p_- $
behaves as
$(p_-)^{{D-4\over 2} + \sum_i |m^1_i-m^3_i|+|m^4_i-m^2_i|}$
where we set $M=0$.
(When $M\neq 0$, one gets an extra power of $p_-$ 
from the propagator.)
Thus as claimed, one could have a singular
contribution $1/\sqrt{p_-}$ for $D=3$.

To check finiteness of the sum over $m_i$, 
let us look at large $m_i$ behaviors. When $\beta^2 >\epsilon$
with finite $\epsilon$, the series is convergent exponentially
due to the $(1-\beta^2)^{m_i}$ term and 
the corresponding sum is finite in this case.
Thus only potential danger may appear when $\beta^2=p_-/p^1_-$ becomes
very small. Since, the lowest power terms in $p_-$ occur when 
$l^3_i=m^3_i$ we  only focus on these terms in the sum. 
Then
the expression in Eq.(\ref{eventchannel}) becomes
\bea
&& F_{2\vm,2\vm_3;2\vm_1}(p_-,p_-^3;p_-^1)\sim 
\left({\mu p_-\mu p_-^3 \over \pi^2 e^4 \mu  
p^1_-}\right)^{{D-2\over 4}} \prod_{i=1}^{D-2}\Biggl[ 
(-1)^{m_i} ( m_i)^{-1/4}(1-\beta^2)^{m_i}
\cr 
&&\ \ \ \ \ \ \ \ \ \ \ \ \ \ \ \  \ \ \ \ \ \ \ 
\ \ \ \ \ \ \ \ \ \ \ \   
\times
\sum_{l_i=|m_i^1-m_i^3|}^{m^1_i+m_i^3} 
(-1)^{l_i} (2m_i \beta^2)^{l_i} C_{l_i m^1_i m^3_i}
\Biggr]
\eea
where
\bea
C_{l_i m^1_i m^3_i}=
{\sqrt{(2m^1_i)!(2m^3_i)!} \over (l_i+m^1_i-m^3_i)!
(l_i+m^3_i-m^1_i)!(m^1_i+m^3_i-l_i)!
}\,.
\eea
For small $p_-$, one may approximate $\mu/p_-=j_0$ with some large 
integer
$j_0$ and express 
$m_i= k_i j_0+q_i$ with $q_i=0,1,\cdots, j_0-1$.
By summing over $q_i$, one obtains in leading approximation
\bea
 {\cal A}^t_{2\to 2}\sim \prod_{i=1}^{D-2}
\Biggl[ \sum^\infty_{k_i} { k_i^{-3/2}} e^{-({\mu\over p^1_i}
+{\mu\over p^4_i}) k_i
} G_{m^1_im^3_i}(k_i \mu/p^1_i) 
G_{m^4_i m^2_i}(k_i \mu/p^4_i) \Biggr] 
\eea
where the polynomial $G$ is defined by
\bea
G_{m^1_im^3_i}(x) =
\sum_{l_i=|m_i^1-m_i^3|}^{m^1_i+m_i^3} 
 (-2x)^{l_i} C_{l_i m^1_i m^3_i}\,.
\eea
The sum converges and the corresponding contributions are certainly
finite even in the $p_-\to 0$ limit.
For the $u$-channel, the effect is simply exchanging the third particle 
with the fourth particle. Hence their results are essentially
the same as that of $t$-channel.

The four point scattering amplitudes  essentially correspond to  two body
potential between particles as shown in Appendix B.
From this, one may
determine the region of momentum space of 
($p^1_-,p^2_-,p^3_-,p^4_-$) 
corresponding to the 
asymptotic region where particles are separated by large distance.
As shown in the Appendix B, 
the asymptotic
region corresponds to the limit where 
$p_-^3-p_-^1$ or $p_-^4-p_-^1$ become very small while other two independent
combinations of momenta kept fixed. By the Fourier 
transformation of the results (\ref{singular}), 
we conclude that
the potential $V(x^-_1\!-\!x^-_2)$ falls off  faster than or equal to
$1/|x^-_1\!-\!x^-_2|$
for $D > 3$
and 
$1/\sqrt{|x^-_1\!-\!x^-_2|}$
for $D=3$.

Here we like to comment on the strength of scalar interactions
as a function of occupation numbers of the 
transverse harmonic oscillators. In short, for large 
occupation numbers, the interaction is still finite and well 
behaved. To see this, we consider
the characteristic vertices in Fig. 1, whose strength is 
basically  given by
$\lambda  F_{\vn_1,\vn_2;\vn}(p^1_-,p_-^2;p_-)$.
To illustrate the characteristic strength in the large occupation numbers,
let us consider for example the case where $\vn=\vn_1+\vn_2$ and 
$p_-=p_-^1+p_-^2$. The explicit evaluation leads to
\bea
 F_{\vn_1,\vn_2;\vn}(p^1_-,p_-^2;p_-)=\prod_i
\left({\mu p^1_-\mu p_-^2 \over \pi \mu 
p_-}\right)^{{1/ 4}} 
\left(
{ (n_i^1+n_i^2)!\over n_i^1! n_i^2!}\right)^{{1/ 2}}
\left({p^1_-\over p_-}\right)^{{n_i^1/2}}
\left({p^2_-\over p_-}\right)^{{n_i^2/ 2}}
\eea
For fixed $N_i=n_i^1+n_i^2$, the maximum occurs at $p^1_-=p^2_-$
and $n_i^1=n_i^2$, in which $F$ behaves as
\bea
 F\sim \prod_i
\left({\mu p_-\over 2 \pi^3   
}\right)^{{1/ 4}} 
{ 1\over  e\, n_i^{{1/ 4}}}
\eea
Similar trends follow 
for the generic $F$'s including those involved in other
channels; in the limit of  the large occupation numbers 
they are not diverging. 
Hence we conclude that the  strength of interaction with large external
occupation number is also fine.

One may suspect that our results about the existence of S-matrix is strongly 
depending on the form of the $\phi^3$ interactions we have turned on. Had we 
taken the local quartic interaction,
the contribution to the four point amplitudes would behave as
\bea
 {\cal A}\sim
{\lambda'\over \sqrt{p_-^1 p_-^2 p_-^3 p_-^4}}
\int d\vec{z}\,\, K_{\vn_1}K_{\vn_2}K_{\vn_3}K_{\vn_4}
\eea
which is well behaved and finite in any case. 
We would like to note again  that the effective field theory of $\phi_{\vn}$ modes
is not a local theory in the two dimensional sense.
The interactions
of higher than fourth order do not contribute to the
two particle amplitudes in the tree level and lead to similar result for the 
$n_i\ particle\to n_f\ particle$ amplitudes and hence they do not
affect the existence of interaction free states in the asymptotic
regions. Therefore, we may conclude that for any scalar field theory in the plane 
wave background with polynomial potential we have (classically) well-defined 
S-matrix.

\section{Gauge Theory on Plane Waves}

Unlike the scalar field theories, the form of the 
couplings
are  completely fixed in  the gauge theories.
With this strongly constrained form of   interactions, it is 
interesting to ask whether the theory allows the 
asymptotically free states, again in the 1+1 dimensional 
sense.
The gauge theory in the plane wave background is an example
of gauge theories of only massive degrees. The mechanism is 
completely different from the case of spontaneous symmetry 
breaking; rather the mass follows from the geometry. Namely
the mass comes from the confining potentials in the 
transverse directions and  similar to the case of scalar theory, 
the theory effectively becomes a {\it non-local} 1+1 dimensional  theory.

It is well known that 
the Coulomb interactions in 1+1 dimensions do not fall off at large 
distance, so the asymptotically free charged states cannot 
exist.\footnote{In this paper, we consider only the S-matrix formulation 
in which 
 asymptotic  states are formed by
the 
original free states defined by simply shutting off interactions. 
 It is quite possible that one may in general 
 have  asymptotic states that differ from the original 
free states (e.g. this is the case for the confining theories).} 
This is essentially true for any two dimensional massless field theory, such as 
$2D$ QED, or $2D$ gravity.  
Since degrees along the transverse $D-2$ 
directions   are again confining, 
it is a priori not clear whether
the effectively 1+1 dimensional gauge theory are free of such 
confining Coulomb interaction or not. Below we shall investigate
such an issue via the light-cone gauge fixing. Indeed the 
Coulomb interaction is apparently there producing singular 
interactions of  $(1/p_-)^2$ where $p_-$ is the exchanged light-cone
momentum.  However in any tree amplitude one may show that 
all such singular contributions precisely cancel out.

There is another potential danger; on-shell photon exchanges may
appear, causing further problems.
Here we show in detail 
how one gets precise cancellation between 
terms appearing
in the action such that the exchanged photons are always off-shell.

To avoid the unnecessary complications here we only consider the $U(1)$ gauge 
theory coupled to a scalar field:\footnote{In principle one can consider 
fermions. However, in order to formulate fermions in the plane wave 
background, one should consider the form fluxes present in the corresponding 
supergravity solutions. (Such fluxes lead to the fermionic mass terms.) 
Here we will not consider fermions. However, we believe that our result about 
existence of the S-matrix can be extended to those cases as well.}
\be\label{vectorlagrangian}
{\cal L}=-{1\over 4} g^{\mu\alpha}g^{\nu\beta}F_{\mu\nu}F_{\alpha\beta}
-{1\over 2} g^{\mu\nu} (D_{\mu}\phi)^\dagger D_\nu\phi 
-{1\over 2} M^2 \phi^\dagger\phi\ ,
\ee
where $g^{\mu\nu}$ is the inverse of the metric (\ref{wave}), i.e.
\[
g^{+-}=g^{-+}=-1\ , \ \ \ g^{++}=0\ , \ \ \ g^{--}=\mu^2 \sum_{i=1}^{D-2} 
z_i^2\ ,\ \ \ g^{ij}=\delta^{ij}\ ,
\]
$D_\mu \phi=\partial_\mu \phi -ie A_\mu \phi$ and $F_{\mu\nu}=\partial_\mu 
A_\nu-\partial_\nu A_\mu$.

\subsection{Fixing the light-cone gauge}

In order to study any physical processes such as scattering, one needs to fix 
a gauge. Because of the form of our background it turns out to be more 
appropriate to fix the light-cone gauge
\be\label{gauge}
A_-=0\ .
\ee
As it is usual in the light-cone gauge theory analysis, since the other 
light-cone component $A_+$ is not a dynamical one, one can solve the equation 
of motion for $A_+$ and plug the solution back into the action. In this way 
we find an action which only involves real physical degrees of freedom. 
Moreover, in any axial gauge such as light-cone, ghosts are decoupled 
\cite{lightcone}
and we do not need to worry about them.
Imposing the $A_-=0$ condition the Lagrangian (\ref{vectorlagrangian}) 
simplifies as
\bea\label{veclagrangian}
{\cal L}&=&{1\over 2}(\partial_-A_+)^2-\partial_-A_i \partial_iA_++ e J^+A_+ 
 \cr &&
+\partial_+A_i\partial_-A_i
-{1\over 2} \mu^2 \sum_{i=1}^{D-2} z_i^2 (\partial_-A_j)^2-{1\over 4} F_{ij}^2 
\cr 
&& +{1\over 2}[
(\partial_+\phi)^\dagger\partial_-\phi+(\partial_-\phi)^\dagger\partial_+\phi]
-{1\over 2} \mu^2 \sum_{i=1}^{D-2} z_i^2 (\partial_-\phi)^\dagger 
\partial_-\phi-{1\over 2}
(\partial_i\phi)^\dagger\partial_i\phi-{1\over 2}M^2\phi^\dagger\phi\ \cr
&& - eJ_iA_i-{1\over 2}e^2 \phi^\dagger A_iA_i\phi,
\eea
where
\bea
J^+=
{i\over 2}\left(\phi^\dagger\partial_-\phi
-(\partial_-\phi)^\dagger\phi\right)\,, \ \ 
J_i=\,\,{i\over 
2}\left(\phi^\dagger\partial_i\,\phi-(\partial_i\,\phi)^\dagger\phi\right)
\eea
with $\phi=\phi_1+i\phi_2$.
As we see the only $A_{+}$ dependent part is in the first line of 
the Lagrangian (\ref{veclagrangian}). Then the equation of motion for $A_+$ 
reads as
\[
\partial^2_-A_+=\partial_-\partial_iA_i +e J^+\ .
\]
Since we consider only the field configurations with $p_-\neq 0$, 
$\partial_-^2$ is formally invertible and hence
\be\label{A+}
A_+={1\over \partial_-}\partial_iA_i+ e{1\over \partial_-^2}J^+\ .
\ee 
Inserting the above into the Eq.(\ref{veclagrangian}) we obtain the 
fully gauge-fixed Lagrangian
\bea\label{GFlag}
{\cal L}&=& \partial_+A_i\partial_-A_i
-{1\over 2} \mu^2 (\sum_{i=1}^{D-2} z_i^2) (\partial_-A_j)^2-{1\over 
2} (\partial_iA_j)^2 
\cr 
&&+{1\over 2}[
(\partial_+\phi)^\dagger\partial_-\phi+(\partial_-\phi)^\dagger\partial_+\phi]
-{1\over 2} \mu^2 \sum_{i=1}^{D-2} z_i^2 
(\partial_-\phi)^\dagger\partial_-\phi-{1\over 2}
(\partial_i\phi)^\dagger\partial_i\phi-{1\over 2}M^2\phi^\dagger\phi\ 
\cr
&&-{1\over 2}e^2 {1\over \partial_-}J^+\ {1\over \partial_-}J^+- e
\partial_iA_i{1\over \partial_-}J^+-eJ_iA_i-{1\over 2}e^2 
\phi^\dagger A_iA_i\phi\,.
\eea
To run the perturbation theory machinery we need to solve the equation of 
motion of the quadratic parts of the action. The equations of motion for 
$\phi$ and $ 
A_i$ fields are exactly the same as Eq.(\ref{eom}) (of course for gauge fields 
we should set $M=0$), and hence the solutions for them are similar to 
Eq.(\ref{Phi}). For the gauge field, it is
\be
A^j=\sum_{\vn}{1\over 2\pi}\int 
dq_+\int_0^\infty {\,\,\,dq_- \over 
\sqrt {2 q_-}}
 A^j_{\vn}(q_-,q_+)\prod_{i=1}^{D-2} K_{n_i}({\sqrt{\mu q_-}z_i}) 
e^{-i(q_-x^-+q_+ x^+)}+ {\rm h.\ c.} 
\ee
and  the scalar fields may be expressed as
\be
\phi_a=\sum_{\vn}{1\over 2\pi}\int dq_+\int_0^\infty {\,\,\,dq_- \over 
\sqrt {2 q_-}}
 \phi_{a \vn}(q_-,q_+)\prod_{i=1}^{D-2} K_{n_i}({\sqrt{\mu q_-}z_i}) 
e^{-i(q_-x^-+q_+ x^+)}+ {\rm h.\ c.} 
\ee
with $a=1,2 $ representing the real and imaginary parts of $\phi$.
Then the quadratic part of the action written in terms of Fourier modes is 
obtained to be
\bea
{\cal L}_0 &=&
\sum_{\vn} \phi^{\dagger}_{(+)\vn}(p_-
)\left( p_+ -
{\cal E}(\vn,p_-)
\right) \phi_{(+)\vn}(p_-
)+
\sum_{\vn} \phi^{\dagger}_{(-)\vn}(p_-
)\left( p_+ 
-{\cal E}(\vn,p_-)\right) \phi_{(-)\vn}(p_-
)\cr
&+&\sum_{\vn} A^{j*}_{\vn}(p_-
)\left( p_+ 
-\mu  {\cal N}\right) A^j_{\vn}(p_-
)\ 
\eea
where we have suppressed $p_+$ dependence in the component fields
and introduced
\be
\phi_{(\pm)\vn}= {1\over \sqrt{2}}(\phi_{1\,\vn}\pm
i\phi_{2\,\vn})\,.
\ee
The $(\pm)$ components carry respectively $(\pm e)$ charges.

\begin{figure}[htb]
\vskip 1cm 
\epsfxsize=4.5in
\centerline{
\epsffile{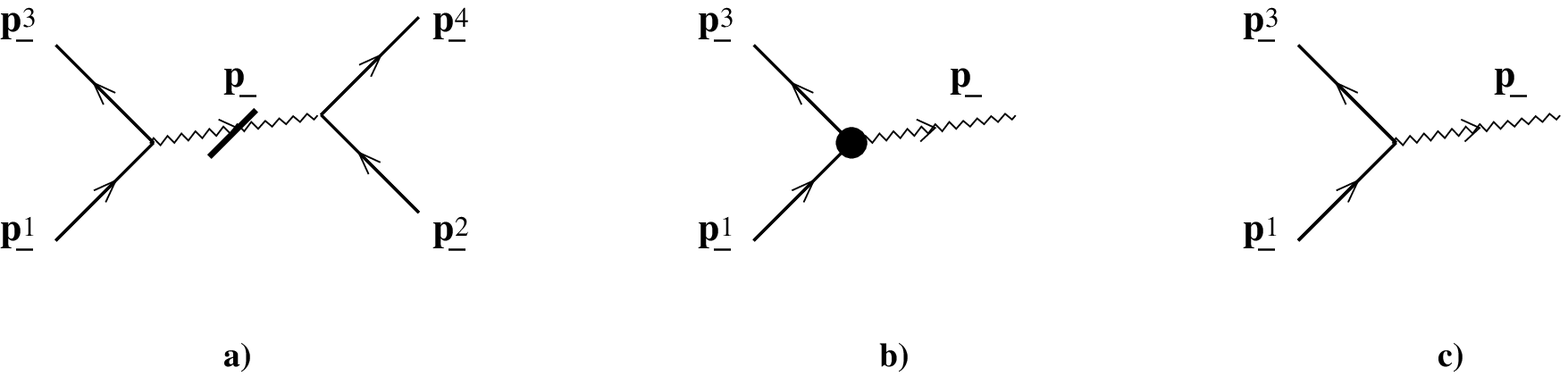}
}
\vspace{.1in}
{\small Figure~3: Vertices for the $(+)$ charges.
\newline
a) represents  ``the Coulomb interaction'', 
${-i e^2 (p^1_-+p^3_- )(p^4_-+p^2_- )
\over (p_-)^2
(2p^1_-
2p^2_-2p^3_- 2p^4_-)^{1/2}}
\int d \vec{x}\,\,
K_{\vec{n}_1} K_{\vec{n}_2}K_{\vec{n}_3}K_{\vec{n}_4}$ with 
$p_-=|p^1_--p^3_-|$. 
b) corresponds to
${ i e 
(p^1_-+p^3_- )\over 
 p_-
(2p^1_-
2p^3_-2p_- )^{1/2}}
\int d \vec{x}\,\,
K_{\vec{n}_1} K_{\vec{n}_3}\partial_i K_{\vec{n}}$.
The 
third diagrams is given by
$-{ i e \over 
(2p^1_-
2p^3_-2p_- )^{1/2}}
\int d \vec{x}\,\,
 K_{\vec{n}}(K_{\vec{n}_1} \partial_i K_{\vec{n}_3}
- K_{\vec{n}_3} \partial_i K_{\vec{n}_1}
)$. These diagrams, a), b) and c) represent the first, second and third 
term of the third line of Eq.(\ref{GFlag}), respectively. There are many more 
diagrams 
differing from above only 
 by orientation and  directions of arrows. Here we do not draw 
all of them.  In case of $(-)$ charges, $(+e)$ should be replaced 
by $(-e)$ accordingly.
}
\vskip .5cm 
\end{figure} 

For our later analysis it is also useful to write $J^+$ in terms of these 
Fourier modes
\bea
J^+&=&\sum_{\vn,\vm}{1\over (2\pi)^2}
\int dp_+dp'_+\int_0^{\infty} {dp_-dp'_-\over
\sqrt{2p_-}\sqrt{2p'_-}}
\,\,(p_-+p'_-)\,\,
\phi_{(+)\vn}(p_-,p_+)\phi^{\dagger}_{(+)\vm}(p'_-,p'_+)\cr &&
\ \ \prod_{i=1}^{D-2} K_{n_i}({\sqrt{\mu p_-}z_i})K_{m_i}({\sqrt{\mu 
p'_-}z_i})
e^{-i(p_--p'_-)x^--i(p_+-p'_+)x^+} - [(+)\rightarrow (-)]\,.
\eea

\subsection{Evaluating $2\phi\to 2\phi$ amplitude}

As in the previous section again we focus our analysis on the $2\phi\to 2\phi$ 
{tree level} scattering processes. As depicted in Fig. 4, 
there are 
five diagrams contributing to this amplitude where  each diagram can be in $s$, 
$t$ or $u$ channel.

\begin{figure}[htb]
\epsfxsize=5.0in
\centerline{
\epsffile{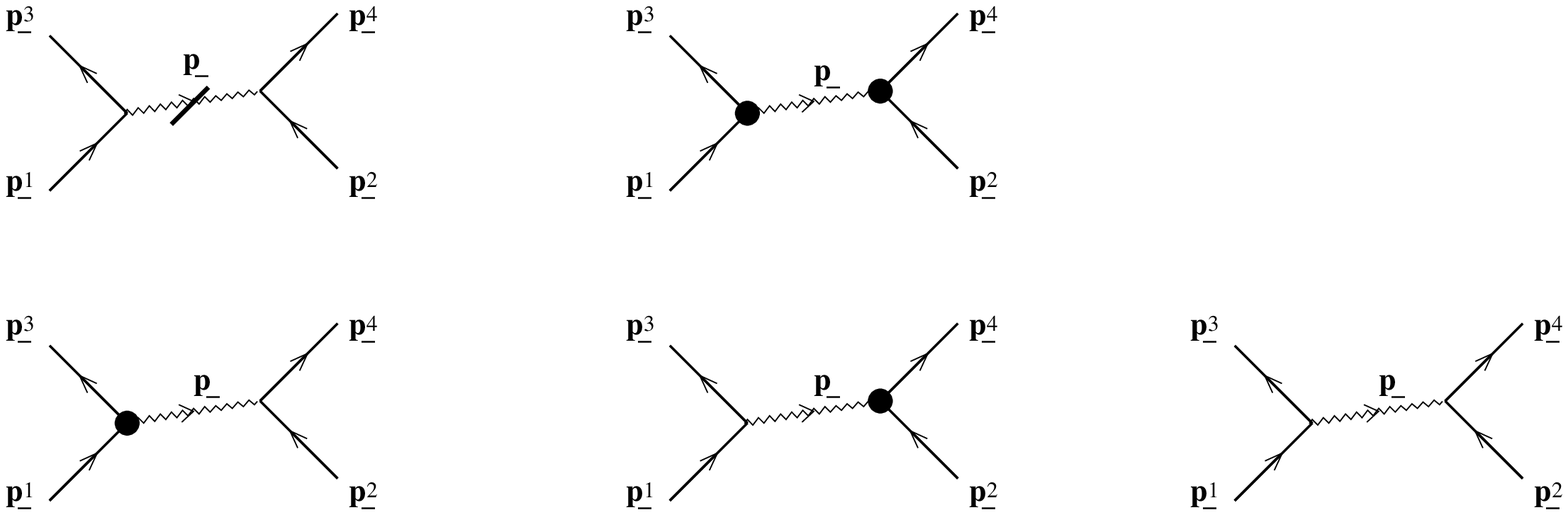}
}
\vspace{.1in}
{\small Figure~4: Four point amplitudes of  $2\phi\to 2\phi$ scattering. 
Here we have depicted  only $t$ channel diagrams. In  the 
$2\phi\to 2\phi$ scattering, there are also $u$ and $s$ channel
contributions which have not been drawn here.  }
\vskip 1cm 
\end{figure}

As we have discussed the question of having a (classically) 
well-defined S-matrix is now equivalent to having finite amplitudes for any 
generic in and out -going state. Similar to the $\phi^3$ case there are 
two different potential dangers. It turns out that the arguments for the 
convergence of the sum over all possible exchanged states 
is quite similar to the $\phi^3$ case, hence we do not repeat them here. 
However as we will see, because of having spatial derivatives in the 
interaction vertices, there is more room for  having on-shell photon 
exchange, which will be considered separately in the next subsection.

It is known that for the usual QED in the light-cone gauge we have 
$\left({1\over p_-}\right)^2$ and ${1\over p_-}$ singularities 
in some of the graphs contributing to 
the $e-e$ scattering. However, summing up all the graphs, as we physically 
expect, these singularities cancel out and we remain with 
a smooth $p_-\to 0$ 
limit at least at the tree level. As mentioned earlier, 
these singular
contributions come from the Coulomb type interaction of the
1+1 dimensions. 
Since the physical flux between charges
spreads over all over $D-1$ dimensions in the case of flat space,
these singular contributions are simply
a light-cone gauge artifact. 
Due to the shape of plane wave geometry and the confining
potential however, such a simple argument of cancellation of the flat 
space does not apply for gauge theory of the plane waves.
Therefore, besides the two issues that are relevant for the plane wave 
background, we first need to clarify the $p_-\to 0$ behavior. In the 
following we show that the same cancellation of $p_-$ poles is also present  
in the plane wave case. Note that in the plane wave case ``effective mass'' is 
also proportional to $p_-$ and hence the $p_-\to 0$ limit should be handled 
with a special care.

To see the cancellation of 
the singular contributions explicitly, we focus on the four point 
interactions of $(+)$ charges depicted in Fig. 4. The first term represents
instantaneous Coulomb interaction in the light cone direction.
It is of order $\left({1\over p_-}\right)^2$ and its explicit 
expression reads as
\be
{\cal A}_a ={-i e^2 W\over (p_-)^2}\int d \vec{x}\,\,
K_{\vec{n}_1} K_{\vec{n}_2}K_{\vec{n}_3}K_{\vec{n}_4}\ ,   
\ee
where we have introduced the kinematical factor $W$:
\be
W\equiv  {(p^1_-+p^3_- )(p^4_-+p^2_- )\over \sqrt{2p^1_-}
\sqrt{2p^2_-} \sqrt{2p^3_-} \sqrt{2p^4_-}}\,.  
\ee
The second diagram is  also of order $\left({1 \over p_-}\right)^2$ and 
expressed as
\be
{\cal A}_b ={-i e^2 W\over (p_-)^2}\sum_{\vec{n}}\int d \vec{x}\,\,
\partial_i\left(K_{\vec{n}_1} K_{\vec{n}_3}\right) K_{\vec{n}}\,\, 
{1\over 2p_-(p_+-{\cal E}(\vec{n},p_-))} 
\int d \vec{y}\,\,
K_{\vec{n}}
 \partial_i\left(K_{\vec{n}_4} K_{\vec{n}_2}\right)\,.
\ee
Now note  the identity
\bea
&&\sum_{\vec{n}} K_{\vec{n}}(\sqrt{\mu p_-}\vec{x})\,\, 
{1\over 2 p_-(p_+-{\cal E}(\vec{n},p_-))} 
K_{\vec{n}}(\sqrt{\mu p_-}\vec{y})
=\langle \vec{x}|\,\, 
{1\over 2 p_-p_+ 
+\nabla^2 -(\mu p_-\vec{y})^2} |\vec{y}\rangle
\,\cr
&&=\langle \vec{x}|\,\, 
{1\over \nabla^2} |\vec{y}\rangle
-2p_-p_+ \langle \vec{x}|\,\, 
{1\over (\nabla^2)^2} |\vec{y}\rangle + O[ (p_-)^2]\,,
\label{identity}
\eea
where we have made expansion with respect to $p_-$ in the second line.
Using this identity, after some algebraic manipulations 
one can  show that
\bea
{\cal A}_b ={i e^2 W\over (p_-)^2}\Biggl(&&
\int d \vec{x}\,
K_{\vec{n}_1} K_{\vec{n}_3} K_{\vec{n}_4} K_{\vec{n}_2}\cr
&&
- 2p_-p_+ 
\int d \vec{x} d \vec{y}\,
K_{\vec{n}_1} K_{\vec{n}_3}\,
\langle \vec{x}|\,\, 
{1\over \nabla^2} |\vec{y}\rangle\,
K_{\vec{n}_4} K_{\vec{n}_2} +  O[ (p_-)^2]
\Biggr)\,.
\eea
The first term in the parenthesis is  precisely canceling the
contribution of ${\cal A}_a$. In a similar manner, one may show that
the second term is canceling with the leading contributions
 of the third and the fourth diagrams of
Fig. 4, which are of order $1/p_-$. Therefore we are  left only 
with terms that are non-negative powers of $p_-$.

The cancellation actually occurs in all possible tree diagrams 
of the gauge theory. In this cancellation, the fact that
the effective confining potential in the transverse directions
is proportional to $(p_-)^2$  plays an important role.
Because of this, the effective transverse potential 
for the exchanged  photon
only contributes
to the order of  $(p_-)^2$ in the formula (\ref{identity})
and does not affect at all the cancellation of the singular
interactions.
If the effective transverse potential 
were too stiff, the Coulomb
interaction would become essentially one dimensional and
no such cancellation would occur. (With stiff enough transverse potential,
 fluxes cannot leak out 
in the transverse direction,  producing one dimensional 
confining potential between 
charges.)
Fortunately the potential
is proportional to $(p_-)^2$ and becomes weak enough in the limit 
$p_-\rightarrow 0$ which corresponds to the region of the
large separation of charges.

\subsubsection{On-shell photon exchange is not possible}

Similar to the $\phi^3$ case, kinematics cannot prevent us from having 
on-shell photon exchange. Therefore the only way to save the theory is that 
the interactions are zero exactly when the parameters of the external 
particles allow the on-shell photon exchange.
Dealing with a vector particle, $\phi\phi\gamma$ interaction terms involve 
space derivative of the fields and in this respect this case is different from 
the $\phi^3$ case.
In particular, noting the Eq.(\ref{derivative}) of the appendix and 
Eq.(\ref{sf}) or 
(\ref{sf1}) it seems quite possible to have on-shell propagation of exchanged 
photon in the $D=4,\ M=0$ case (this will become apparent momentarily). However, 
noting Eq.(\ref{sf}) we see that still $D>4$ is  safe 
from this potential danger. 
The above argument can be repeated for spin $S$ particles. In general the 
coupling of spin $S$ particle to scalar fields involves $S$ number of 
derivatives and according to our arguments $D\leq 2S+2 $ may be problematic. 
For example, for scalars coupled to six dimensional gravity 
$(S=2)$ on-shell propagation of exchanged gravitons is kinematically allowed.
In what follows we present explicit calculations showing that the interaction
exactly for the on-shell exchange of photon turns off, removing the potential 
danger. 

For this, we consider 
the terms in the interaction 
Lagrangian which may cause a problem, i.e. those which involve spatial 
derivatives:
\be
S_{int, \partial}=-e\int dx^+dx^-d^{D-2}z \left(\partial_i A_i\ {1\over 
\partial_-}J^++J_iA_i\right)\ .
\ee
Since we want to check whether for  certain modes the above term is vanishing 
or not, in our computations we will drop the overall factors;   
\bea\label{danger}
S_{int,\partial}\sim \int d\vec{z}
&\Biggl(&
{p_-+p'_-\over q_-}\partial_i K_{\vm}(\sqrt{\mu q_-}z_i)
K_{\vn}(\sqrt{\mu p_-}z_i)
K_{\vn'}(\sqrt{\mu p'_-}z_i)
\cr
&&+
K_{\vm}(\sqrt{\mu q_-}z_i) 
\partial_i K_{\vn}(\sqrt{\mu p_-}z_i)
K_{\vn'}(\sqrt{\mu p'_-}z_i)
\cr
&& -K_{\vm}(\sqrt{\mu q_-}z_i)
K_{\vn}(\sqrt{\mu p_-}z_i)
\partial_i K_{\vn'}(\sqrt{\mu p'_-}z_i)
\Biggr)\ .
\eea
Note that the above expression is the one appearing in the $t$ or $u$ -channel 
where $q_-$ is the light-cone momentum of photon and, $p_-$ and $p'_-$ those of 
in and out -going scalars, therefore $p_-=p'_-+q_-$.

We are only interested in the cases where all three particles can be on-shell 
at the same time. This implies that $q_+=\sum_i m_i +{D-2\over 2}$. Since 
$p_+=p'_++q_+$ due to energy conservation, for the cases of our interest
(which may cause a potential danger)
\be\label{3onshell}
\sum_{i=1}^{D-2} n_i-(n'_i+m_i)={D-2\over 2} \ .
\ee
On the other hand as we discussed in the previous section, the integral
\[
\int H_n(z) H_{n'}(\alpha z) H_m(\beta z) e^{-z^2} dz
\]
is non-zero only for $n\leq n'+m$. Therefore, having a single derivative on 
$K_{n_i}$'s the only modes which contribute to the interactions are those with
$\sum_i n_i-(n'_i+m_i)\leq 1$. Eq.(\ref{3onshell}) and this condition can be 
satisfied simultaneously only for $D=4$ and $\sum_i n_i-(n'_i+m_i)=1$. So, we 
evaluate the integral (\ref{danger}) only for the cases where one of $n_i$'s 
(say  $n_1$) is $n'_1+m_1+1$ and for the rest $n_i=n'_i+m_i$, i.e.
\bea\label{danger1}
S_{int,\partial}^{relevant}\sim 
\int d\vec{z} &&\!\!\!\!\!\!
\bigg[\partial_i K_{\vm}(\sqrt{\mu q_-}z_i) 
\left({p_-+p'_-\over 
q_-}+1\right) K_{\vn}(\sqrt{\mu p_-}z_i)
K_{\vn'}(\sqrt{\mu p'_-}z_i)\cr
&&+
2K_{\vm}(\sqrt{\mu q_-}z_i) 
\partial_i K_{\vn}(\sqrt{\mu p_-}z_i)
K_{\vn'}(\sqrt{\mu p'_-}z_i)
\bigg]_{\sum_i n_i-(n'_i+m_i)=1}
\eea
Using Eqs.(\ref{harmonic}), (\ref{special}) and (\ref{derivative}) 
 after some 
algebra we find
\be\label{danger2}
S_{int,\partial}\mid_{\sum_i n_i-(n'_i+m_i)=1}\sim 
\left[\sqrt{q_-}\sqrt{{q_-\over p_-}} \left(1+{p_-+p'_-\over q_-}
\right)N_{n_i}^{-2}
-4 \sqrt{p_-} n_i N_{n_i-1}^{-2}\right]\ ,
\ee
where $N_n$ is the normalization factors defined in (\ref{norm}). Note that 
to arrive at Eq.(\ref{danger2}) we have dropped all the normalization 
and overall factors. Inserting the value of $N_{n_i}$'s it is readily seen 
that $S_{int,\partial}^{relevant}=0$, that is, the exchanged photon 
is never on-shell. All the above manipulation can be repeated for the 
$s$-channel exchanges, leading to the same result. Hence, the $2\phi\to 2\phi$ 
amplitude (at tree level) is finite, with a smooth $p_-$ dependence. It is 
straightforward to check $\phi\gamma\to \phi\gamma$ scattering and observe 
that it is well-behaved scattering as well. 
Therefore the special form of the interactions which is fixed by the gauge 
symmetry guarantees that we have a notion of well-defined asymptotic states, 
and therefore S-matrix, for gauge theories on the plane wave background for 
any $D>2$.

\section{Discussion and Remarks}

In this work  the question of existence of 
S-matrix for string/field theories on 
the plane wave background has been addressed. A priori there are several 
pros and cons, and hence the question asks for a 
direct and explicit 
analysis.
Considering scalar field theories on the $D$ dimensional 
plane wave background we 
argued that, because of the specific form of the geometry the field theory
can effectively be considered as a {\it non-local} $1+1$ dimensional field 
theory for the 
``harmonic oscillator'' modes coming from the state of the field in the $D-2$ 
dimensional transverse directions. As we showed this is a general 
feature of any 
field theory on this background. Through explicit calculation of a 
generic four 
point function at tree level, we showed that these four point functions are 
smooth enough to allow an S-matrix interpretation. Hence we concluded that we 
have S-matrix formulation. Although we have not checked explicitly, we believe 
that the above result can be extended to the case of gravity (spin two 
particles).

In course of the computations in the gauge theory case (section 3.2.1) 
we found a 
non-trivial and amazing cancellation in the $\phi\phi\gamma$ 
interaction vertices 
for exactly the modes that can lead to on-shell photon exchange. 
Presumably
this cancellation is a consequence of the gauge symmetry. 
It would be interesting 
to show explicitly how gauge symmetry is responsible for 
this cancellation.
 
In this work we only focused on the tree level checks of the existence of 
the S-matrix. However, generally non-renormalizability of the theory may
spoil 
the notion of unitary S-matrix 
at quantum (loops) level. 
In the field theories we have studied, we had in mind that they are 
coming as effective dynamics of strings in the plane wave background.
Therefore the renormalizability is not an essential issue
 and should be addressed in the string theory level. 
However, one can think of these field theories independently of 
string theories.
It would then be interesting to check whether 
our statement about the existence  
of S-matrix also holds at quantum level specifically for $D <5$.

With the knowledge of existence of the S-matrix, 
the next step one may pursue
will be the explicit construction of vertex operators for string 
theories in
the plane wave background. With the vertex operators, one can in principle 
compute all the amplitudes using the path integration over 
Riemann surfaces
with insertion of vertex operators.  
As it is known, for the flat background case working in the light-cone gauge 
and for $p_-=0$, the tree and one loop diagrams may be handled without much 
complication using the operator formulation. However, for $p_-\neq 0$ there 
are severe ordering problems \cite{GSW}. In the plane wave background, 
since the nature of physics for $p_-=0$ and $p_-\neq 0$ are completely different, 
we are forced to set $p_-\neq 0$ and hence
these ordering problems should show up in the vertex operators in the 
light-cone gauge, making the formulation of vertex operators more non-trivial.
This will be the primary direction for the further studies.

{\it \bf Note Added:}\footnote{ We would like to thank Juan Maldacena and 
Joe Polchinski for 
discussions on this point.}

In order to discuss the existence of S-matrix for any field theory,
 besides the 
finite-range-force condition that we analyzed in detail here, there 
is another condition: we should 
be able to make ``moving'' wave-packets; after all  S-matrices are 
constructed to encode the 
information regarding the scattering of wave-packets. The latter 
condition, which we had missed in 
the earlier version of this work, seems to be nontrivial for the 
massless ($m^2=0$) case, noting 
the unusual form of the dispersion relation in the plane-wave 
background (\ref{dispersion}). 
Explicitly, once we set $m=0$, the light-cone energy $p_+$ becomes 
completely
independent of the 
light-cone momentum $p_-$ and therefore the group velocity of the 
corresponding wave-packet
$\partial p_+/ \partial p_-$ is zero.  That is, for the massless 
case we cannot kick the waves, and 
they would stay in the position they are created. This subtlety may 
shatter our discussion of the 
existence of S-matrix for the massless case.

One should note that the dispersion relation (\ref{dispersion}) is only 
obtained when interactions 
are totally turned off. However, it is conceivable that the interaction, 
through loop corrections 
or possibly some non-perturbative effects, induces $p_-$ dependent 
modifications in the dispersion 
relation and hence provides the kick for the wave-packet to move. 
Of course in order to check this one needs  to extend the analysis  beyond 
the tree level perturbative ones  presented here.
This point needs a more detailed and thorough study and 
 we defer it to future works.

The significance of the massless case becomes clearer noting 
that all supergravity modes in the 
plane-wave background, which are nothing but the lowest lying modes of 
strings in the same 
background, fall into this category \cite{MT}. As it has been listed 
in \cite{MT} the light-cone 
energy of all these modes is $p_-$ independent and hence the above 
discussions apply to them. 
For these modes, however, one can argue that supergravity 
interaction terms  cannot provide
the possible kick mentioned in the previous paragraph. This 
is because the light-cone mass of these 
particles are protected by supersymmetry (they are BPS) and one 
expects these modes to have the 
same mass in the free and interacting theories 
even non-perturbatively.

Unfortunately, it seems that the dual BMN gauge theory 
has nothing more to offer on this issue. 
This can be seen from the fact that mixing between the single, 
double and in general multi trace 
operators corresponding to string vacuum or supergravity modes 
cannot be fixed using the usual 
re-diagonalization argument of Refs. \cite{mixing1, mixing2}. 
For a more detailed gauge theory 
calculation of this point we refer the reader to \cite{mixing1}.

\vfill
\eject
{\large{\bf Acknowledgements}}

We would like to thank Allan Adams, Keshav Dasgupta, 
Mathew Headrick, Simeon Hellerman, Xiao Liu, Juan Maldacena, Michael Peskin, 
Joe Polchinski, Stephen 
Shenker, Sang-Jin Sin, Mark Van Raamsdonk and especially Lenny Susskind 
for helpful discussions.
The work of M. M. Sh-J. is supported in part by NSF grant
PHY-9870115 and in part by funds from the Stanford Institute for Theoretical 
Physics. The work of D. B. is supported in part by
KOSEF 1998 Interdisciplinary Research Grant 98-07-02-07-01-5 
and by UOS 2002 Academic Research Grant.

\appendix

\section{Some useful identities about Hermite functions}
The harmonic oscillator wave functions are related to the 
Hermite polynomials $H_n (x)$ by
\be\label{harmonic}
K_n(\sqrt{\mu p_-} z)=N_n e^{-\mu p_- z^2/2} H_n(\sqrt{\mu p_-} z)\ ,
\ee
where
\be\label{norm}
N_n^2={1\over 2^n n!} \sqrt{{\mu p_-\over \pi}}\ . 
\ee
Using the identity \cite{Handbook}
\[
\int e^{-x^2} H_{n+2m}(\alpha x) H_n(x) dx= \sqrt{\pi} 
{2^n(2m+n)! \over m!}(\alpha^2-1)^m \alpha^n\ ,
\]
and the orthogonality of $H_n$'s, we have
\be\label{expand}
H_m(\alpha z)=\sum_{k=0}^m A_{km}(\alpha) H_k(z)\ ,
\ee
where $A_{km}(\alpha)$ is 
\be\label{a_k}
A_{km}(\alpha)=\left\{\begin{array}{cc}
{m! \over k!({m-k\over 2})!}(\alpha^2-1)^{{m-k\over 2}} \alpha^k\ \ \ \ 
m-k={\rm even}, \\
0 \ \ \ \ \ \ \ \ \ m-k={\rm odd}
\end{array}
\right.
\ee
Given (\ref{expand}) and using (7.375) of Ref.\cite{Handbook}, 
one obtains,
\be\label{3hermit}
\int e^{-z^2} H_{n}(z) H_m(\alpha z) H_l(\beta z) dz= 
\sum_{p=0}^m\sum_{q=0}^l A_{pm}(\alpha) A_{ql}(\beta) 
{2^s \sqrt{\pi} n!p!q!\over (s-n)!(s-p)!(s-q)!}
\ee
where $2s=n+p+q$ and the integral is non-zero only for integer $s$ with 
$s\geq max\{n,p,q\}$.
A useful special case of the above integral is when $n=m+l$ where we have
\be\label{special}
\int e^{-z^2} H_{m+l}(z) H_m(\alpha z) H_{l}(\beta z) dz=\alpha^m 
\beta^{l}(m+l)! 2^{m+l}\sqrt{\pi}\ .
\ee

The other useful identify which has been used in the calculations of section 
3, is concerning the derivative of the harmonic oscillator wave functions is
$H'_n(x)=2n H_{n-1}(x)$ which implies that
\be\label{derivative}
{d\over dz} K_{n}(z)=\sqrt{{n\over 2}}K_{n-1}-\sqrt{{n+1\over 2}}K_{n+1}\ .
\ee

\section{General nonrelativistic field theory and asymptotic regions}

The 1+1 dimensional theories of our interest has the structure
of nonrelativistic field theory with multi flavor. The symplectic
structure is that of field theories. The four point interaction
preserves particle numbers and there is a translational 
invariance in $x^-$ direction. (Below for the  simplicity,
we shall omit the subscript $-$ in $x^-$ or $p_-$.)
Here we like to consider
most general such nonrelativistic field theories with
quartic interactions and derive two body Schr\"odinger equations
from it in the position space. The potential is closely related
to the four point scattering amplitude and, from this, one may
determine which region of momentum space corresponds to the 
asymptotic region where particles are separated by large distance.
In short, in terms of momenta in the four point amplitudes, 
the asymptotic
region corresponds to the limit where 
$p^1-p^3$ or $p^1-p^4$ becomes very small while other two independent
combinations of momenta kept fixed.

The bosonic field theories of four point interactions is described by
the following Lagrangian
\be
L=i\phi_a^\dagger \dot\phi_a -
\phi^\dagger_a  H_0 \phi_a
-{\lambda^2\over 2}\int dx_i\,\,
\phi^\dagger_a (x_1) \phi^\dagger_b (x_2)
V_{abcd}(x_1,x_2,x_3,x_4)
\phi_c (x_3)
\phi_a (x_4)\,,
\ee 
where we assume the free Hamiltonian $H_0$ is diagonalized by
the momentum eigenstates. Without loss of generality, one may take the 
potential satisfying the exchange symmetries,
\bea
 V_{abcd}(x_1,x_2,x_3,x_4)=V_{bacd}(x_2,x_1,x_3,x_4)
= V_{abdc}(x_1,x_2,x_4,x_3)\,.
\eea 
Due to the translational symmetry in $x$ direction, the potential 
only depends upon differences of coordinates and may be written as
\bea
&&V_{abcd}(x_1,x_2,x_3,x_4)=V_{abcd}\left(x_1-x_2,x_3-x_4,
{x_1+x_2\over 2}-{x_3+x_4\over 2}\right)\cr
&&=\int dq dq' dQ \,\,\tilde{V}_{abcd}(q,q',Q)
\, e^{iq(x_1-x_2)}e^{-iq'(x_3-x_4)}
e^{i{Q\over 2}\left(x_1+x_2-x_3-x_4\right)}\,.
\eea 
The tree-level four point amplitude is then given by
\bea
{\cal A}_{ab\rightarrow cd}\sim \lambda^2 
\tilde{V}_{abcd}\left({k_1-k_2\over 2},{k_3-k_4\over 2},k_1+k_2\right)
\,.
\eea 

Upon quantization, the equal time commutation relations are given by
\be
[\phi_a (x),\phi^\dagger_b (x')]=\delta_{ab}\delta (x-x')\,.
\ee
Using the two body wave function defined by
\be
\Phi_{ab}(x_1,x_2)\equiv \langle 0|\phi_a(x_1,t)\phi_b (x_2,t) |\Psi \rangle\,,
\ee
and the operator Schr\"odinger equation $i\dot \phi (x,t)=[\phi(x,t), H]$,
one may get the two body Schr\"odinger equation,
\be
i\dot\Phi_{ab}(x_1,x_2)=
H_0 \Phi_{ab}(x_1,x_2) +
\lambda^2 \int  dx_3 dx_4
V_{abcd}(x_1,x_2,x_3,x_4) \Phi_{cd}(x_3,x_4)\,.
\ee
In the
eigenbasis of total momentum 
$\Phi_{ab}(x_1-x_2, Q)e^{i {Q\over 2}(x_1+x_2)}$, the
Schr\"odinger equation becomes
\be
i\dot\Phi_{ab}(x_r,Q)=
H_0 \Phi_{ab}(x_r,Q) +
\lambda^2(2\pi)^2 \int  dq dq'\,\, e^{iq x_r}\,
\tilde{V}_{abcd}(q,q',Q) \Phi_{cd}(q',Q) \,,
\ee
where $x_r=x_1-x_2$ and
\bea
\Phi_{ab}(q,Q) 
={1\over 2\pi}\int dx_r  \Phi_{ab}(x_r,Q) 
\,e^{-iqx_r}\,.
\eea 
Let us further introduce $\bar{V}_{abcd}(y,y',Q)$ by
\bea
\tilde{V}_{abcd}(q,q',Q)
={1\over 8\pi^2}
\int dy dy'  \,\,\bar{V}_{abcd}(y,y',Q)
\, e^{-{iy\over 2}(q-q')}e^{-i{y'\over 2}(q+q')}\,.
\eea 
It is then straightforward to check that the two body Schr\"odinger 
equation becomes
\bea
(i\partial_t-H_0)\Phi_{ab}(x_r,Q) &=&
2\pi\lambda^2 \int  dy'\,
\bar{V}_{abcd}(2 x_r-y',y',Q) \Phi_{cd}(x_r-y',Q) \cr
&=& 2\pi\lambda^2 \int  dy\,
\bar{V}_{abcd}(y,\,2 x_r-y\,,Q) \Phi_{cd}(y-x_r\,,Q)
\,,
\eea
From this expression, it is clear that the large separation
between particles ($x_r\rightarrow \infty$) corresponds to
the region where $y$ or $y'$ of the potential 
becomes large. This in turn implies that $q-q'=k_1-k_3$ or $q+q'=k_1-k_4$
becomes small where we have used the momentum conservation. 
Hence the asymptotic region corresponds to the momentum space region
of four point amplitudes in which  $k_1-k_3$ or $k_1-k_4$ become 
small.


\begin{thebibliography}{99}
\bibitem{Penrose}
R. Penrose, ``Any Spacetime Has a Plane Wave Limit,''Differential Geometry 
and Relativity, Reidel, Dordrecht, 1976, pp. 271.

R. Gueven, ``Plane Wave Limits and T-Duality,''
{\it Phys.Lett.}{\bf B482} (2000) 255, hep-th/0005061.

\bibitem{BMN}

D. Berenstein, J. Maldacena, H. Nastase, ``Strings 
in flat space and pp waves from ${\cal N}=4$ Super Yang Mills,''
{\it JHEP} (2002) {\bf 0204} (2002), hep-th/0202021. 
 


\bibitem{SYM}
C. Kristjansen, J. Plefka, G. W. Semenoff, M. Staudacher, 
``A New Double-Scaling Limit of N=4 Super Yang-Mills Theory and PP-Wave 
Strings,'' {\it Nucl.Phys.} {\bf B643} (2002) 3, hep-th/0205033.

D. J. Gross, A. Mikhailov, R. Roiban, ``Operators with large $R$ charge in 
$N=4$ Yang-Mills theory,''{\it Annals Phys.} {\bf 301} (2002) 31, 
hep-th/0205066.

N. R. Constable, D. Z. Freedman, M. Headrick, S. Minwalla, L. Motl, A. 
Postnikov, W. Skiba, 
``PP-wave string interactions from perturbative Yang-Mills theory,''
{\it JHEP} {\bf 0207} (2002) 017, hep-th/0205089.

A. Santambrogio, D. Zanon, ``Exact anomalous dimensions of {\cal N}=4 
Yang-Mills operators with large 
$R$ charge,'' {\it Phys.Lett.}{\bf B545} (2002) 425, hep-th/0206079.

\bibitem{mixing1}

N. Beisert, C. Kristjansen, J. Plefka, G.W. Semenoff, M. Staudacher,
``BMN Correlators and Operator Mixing in N=4 Super Yang-Mills Theory,''
hep-th/0208178.
     
\bibitem{mixing2}

D. J. Gross, A. Mikhailov, R. Roiban,
``A Calculation of the plane wave string Hamiltonian from $N=4$ 
super-Yang-Mills theory,'' hep-th/0208231.

N. R. Constable, D. Z. Freedman, M. Headrick, S. Minwalla,
``Operator Mixing and the BMN Correspondence,''hep-th/0209002.

\bibitem{SFT1}
M. Spradlin, A. Volovich, ``Superstring Interactions in a pp-wave 
Background,'' hep-th/0204146.

     
\bibitem{SFT2}
M-x. Huang, ``Three point functions of N=4 Super Yang Mills from light cone 
string field theory in pp-wave,''
{\it Phys.Lett.}{\bf B542} (2002) 255, hep-th/0205311.

M. Spradlin, A. Volovich, ``Superstring Interactions in a pp-wave 
Background II,'' hep-th/0206073.

I. R. Klebanov, M. Spradlin, A. Volovich, `` New Effects in Gauge Theory from 
pp-wave Superstrings,''hep-th/0206221.

J. H. Schwarz, ``Comments on Superstring Interactions in a Plane-Wave 
Background,''{\it JHEP} {\bf 0209} (2002) 058, hep-th/0208179.

A. Pankiewicz, ``More comments on superstring interactions in the pp-wave 
background,'' {\it JHEP} {\bf 0209} (2002) 056, hep-th/0208209.

A. Pankiewicz, B. Stefanski, jr., ``PP-Wave Light-Cone Superstring Field 
Theory,'' hep-th/0210246.


\bibitem{causal}

D. Berenstein, H. Nastase, ``On Lightcone String Field Theory From 
Super Yang-Mills and Holography,''hep-th/0205048. 

D. Marolf, S. F. Ross, 
``Plane Waves: To Infinity And Beyond,''hep-th/0208197.
 
\bibitem{MT}
R.R. Metsaev, A.A. Tseytlin,``Exactly solvable model of superstring in 
Ramond-Ramond plane wave background,''{\it Phys.Rev.} {\bf D65} (2002) 126004,
hep-th/0202109. 

\bibitem{Kim}

Y. Kiem, Y. Kim, S. Lee, J. Park, ``PP-wave/Yang-Mills Correspondence: An 
Explicit Check,'' {\it Nucl.Phys.}{\bf B642} (2002) 389, hep-th/0205279.

\bibitem{Lenny}

For a nice overview on the matter see: 
\newline
L. Susskind, ``The World as a Hologram,''
{\it J.Math.Phys.} {\bf 36} (1995) 6377, hep-th/9409089.

\bibitem{Peskin}
M. Peskin, D. Schroeder, ``Quantum Field Theory'', Chapter 7, Addison-Wesley 
Publishers, (1995).


\bibitem{lightcone}
For a review on the light-cone gauge theories see:

S.~J.~Brodsky, H.~C.~Pauli and S.~S.~Pinsky,
``Quantum chromodynamics and other field theories on the light cone,''
Phys.\ Rept.\  {\bf 301}, 299 (1998), hep-ph/9705477.

G.~Leibbrandt,
``Introduction To Noncovariant Gauges,''
Rev.\ Mod.\ Phys.\  {\bf 59}, 1067 (1987).



\bibitem{GSW}

M. Green, J, Schwarz, E. Witten, ``Superstring Theory,'' Vol. 1, Chapter 7,
Cambridge University Press (1987).


%


%


\bibitem{Handbook}
I. S. Gradshtein, I. M. Ryzhik, ``Table of integrals, series, and products,''
(Academic Press, San Diego, 2000). 


\end{thebibliography}
\end{document}